\documentclass[twocolumn,times]{aastex631}

\shorttitle{Ultra-short period Contact Eclipsing Binary system with extreme mass ratio}
\shortauthors{Papageorgiou et al.}
\begin{document}

\title{A Unique Low-Mass Ratio Contact Eclipsing Binary System under the Period Cut-Off }

\correspondingauthor{Athanasios Papageorgiou}
\email{apapageorgiou@upatras.gr}
 
\author[0000-0002-3039-9257]{Athanasios Papageorgiou}
\affiliation{Department of Physics, University of Patras, 26500, Patra, Greece}

\author{Panagiota-Eleftheria Christopoulou}
\affiliation{Department of Physics, University of Patras, 26500, Patra, Greece}

\author{Eleni Lalounta}
\affiliation{Department of Physics, University of Patras, 26500, Patra, Greece}

\author {C. E. Ferreira Lopes}
\affiliation{Instituto de Astronomía y Ciencias Planetarias, Universidad de
Atacama, Copayapu 485, Copiap\'{o}, Chile}
\affiliation{Instituto Nacional de Pesquisas Espaciais (INPE/MCTI), Av. dos Astronautas, 1758, S\~{a}o Jos\'{e} dos Campos, SP, Brazil}

\author[0000-0001-6003-8877]{M\'arcio Catelan}
\affiliation{Instituto de Astrofísica, Facultad de Física, Pontificia Universidad Católica de Chile, Av. Vicu\~na Mackenna 4860, 7820436 Macul, Santiago, Chile}

\affiliation{Millennium Institute of Astrophysics, Nuncio Monseñor Sotero Sanz 100, Of. 104, 7500000 Providencia, Santiago, Chile}

\affiliation{Centro de Astroingeniería, Facultad de Física, Pontificia Universidad Católica de Chile, Av. Vicu\~na Mackenna 4860, 7820436 Macul, Santiago, Chile}

\author{Andrew J. Drake}
\affiliation{California Institute of Technology, 1200 East California Boulevard, Pasadena, CA 
91225, USA}

\author{Panayiotis Xantzios}
\affiliation{Institute for Astronomy, Astrophysics, Space Applications and Remote Sensing, National Observatory of Athens, 15236, Athens, Greece}

\author{Ioannis Alikakos}
\affiliation{Institute for Astronomy, Astrophysics, Space Applications and Remote Sensing, National Observatory of Athens, 15236, Athens, Greece}





\begin{abstract}
We present a multi-band photometric analysis of CRTS J163819.6+03485, the first low mass ratio (LMR) contact binary system with a period under the contact binary (CB) period limit. The unprecedented combination of mass ratio and period makes this system unique for eclipsing binary (EB) research. Using new multi-band photometric observations, we explored the parameter space of this unique total EB system through a detailed scan in the mass ratio~-- inclination plane and using the \textsc{pikaia} genetic algorithm optimizer. The best set of relative physical parameters and corresponding uncertainties was adopted through Markov Chain Monte Carlo sampling of the parameter space. The resulting mass ratio of the system is $q = 0.16 \pm 0.01$. The absolute parameters were derived by adopting an empirical mass-luminosity relation. Period changes are also investigated by using new observations and archival photometric light curves from massive astronomical surveys,  
which revealed in a preliminary solution the presence of a possible low-mass tertiary companion. The origin and evolutionary status of the system are investigated through the detached-binary formation scenario.

\end{abstract}



\section{Introduction} \label{sec:intro}
In the era of large sky surveys where a plethora of new contact binaries are discovered and analyzed through automated processes, ultrashort-period and low-mass ratio systems remain two of the most enigmatic classes that challenge current theoretical models. Ultrashort-period contact binaries (USPCBs) remain important as they allow investigating the evolution of low-mass contact binary stars under the well-known orbital period cutoff, located at approximately 0.22~days \citep{1992Rucinski} or slightly lower ($\sim 0.15- 0.20$~d) \citep{Li2019}. Low-mass ratio (LMRs) contact binaries ($q\leq 0.25$) are of particular interest as the process of exchanging mass and angular momentum between the components can dramatically alter the evolution of both stars, giving rise to tidal instabilities, and finally leading them to coalesce. 
Our review of LMRs \citep{2022MNRAS.512.1244C} with a spectroscopic determination of mass ratio or/and total eclipses revealed that the majority have periods around 0.4 d. With the same criteria, our compilation of USPCBs revealed a concentration of mass ratio of around 0.4 \citep{2023AJ....165...80P}.

CRTS$\_$J163819.6+034852, with coordinates  
RA(J2000)~= 16\,38\,19.67, Dec(J2000)~= +03\,48\,51.64 and  
a period of 0.205332~d, is one of the shortest-period contact binaries found in the Catalina Sky Survey \citep{adea14}. From an analysis of All Sky Automated Survey for supernovae observations \citep[ASAS-SN;][]{2014ApJ...788...48S}, \cite{2018MNRAS.477.3145J} reported a period of 0.2053318~d, a V-band mean magnitude of 14.51~mag, and an amplitude of 0.39~mag. \cite{2022MNRAS.512.1244C} found that it is also an extreme LMR system, making this system one of a kind since it is, to our knowledge, the only one with this rare combination of period and mass ratio.

Here we present the first multi-band follow-up photometric observations of this unique binary (hereafter CRTS$\_$J163819) in combination with archival light curves (LCs) from other surveys, carrying out a period analysis spanning 15 years, with the intent to clarify its nature. 

\section{Observations} \label{sec:Observations}
\subsection{New photometric observations} \label{subsec:NewPhot}
We  observed CRTS$\_$J163819 in two observing runs in 2018 and 2021 with the 2.3 m Ritchey-Chr\'{e}tien Aristarchos telescope at Helmos Observatory, Greece. In the first run, on July 3  2018 as our primary goal was to confirm its  short period, we observed CRTS$\_$J163819 using a  broad VR filter with the RISE-2 CCD camera. RISE-2 has a CCD size of 1k×1k, with a pixel scale of $0.51\arcsec$ and a field of view of $9\arcmin \times 9 \arcmin$. For the second run, on June 25, 26 and July 2 and 3 2021, we observed in $B,V,R,I$ for 180\,s, 90\,s, 40-50\,s, and 60-70\,s, respectively, with a liquid nitrogen cooled Princeton Instruments VersArray CCD camera. This CCD camera has 1024 × 1024 pixels  with  an effective field of view of $4.8 \arcmin$ × $4.8 \arcmin$. 
  
For the preprocessing of raw images (bias subtraction, flat-fielding) and the aperture photometry, we used our fully automated pipeline \citep{2015ASPC..496..181P} that incorporates PyRAF \citep{2012ascl.soft07011S} and the Astrometry.net packages \citep{2010AJ....139.1782L}.
The differential LCs were generated after choosing suitable comparison and check stars close to the target star in the field. These are  2MASS J16382385+0350508 ($J=13.532\pm 0.029$ mag, $H=13.175\pm 0.022$ mag, $K=13.110\pm 0.035$ mag) and 2MASS J1638176+0349143 ($J=12.302\pm 0.024$ mag, $H=12.033\pm 0.027$ mag, $K=11.999\pm 0.026$ mag), respectively. 
The typical errors in the final differential magnitudes throughout the observing run are $8-10$~mmag. 

\subsection{Photometric Observations from Astronomical Sky Surveys} \label{subsec:ArchPhot}
Light curves of CRTS$\_$J163819 were also found in ASAS-SN \citep[in the $V$ filter;][]{2014ApJ...788...48S,2018MNRAS.477.3145J}, Catalina Sky Surveys \citep[CSS;][]{2014yCat..22130009D},
Zwicky Transient Facility \citep[ZTF$\footnote{\url{https://irsa.ipac.caltech.edu/}}$, in $g$ and $r$;][]{2019PASP..131a8003M}, and the Asteroid Terrestrial-impact Last Alert System \citep[ATLAS$\footnote{\url{https://atlas.fallingstar.com/}}$, $o$ and $c$ bands;][]{2018AJ....156..241H} variable star catalogue. 

\begin{deluxetable*}{LCCLC}[!htb]
\tablecaption{The system parameters and uncertainties for CRTS$\_$J163819.6+034852 \label{tab:Tab1}}
\tablewidth{0pt}
\tablehead{
\colhead{}&\multicolumn2c{Physical Parameters}& \multicolumn2c{Approximate Absolute Parameters}\\[-0.3cm]
\colhead{} & \colhead{$q-i$ scan} & \colhead{MCMC} & \colhead{} & \colhead{}
}
\startdata
$q=\frac{M_{2}}{M_{1}}$	    		&			0.16	\pm	0.02  	 		&	0.16	^{+	0.01	}_{-	0.01	}	&		T_{\rm sys} (K)		        		&	6662	\pm	200	\\
T_{r}=\frac{T_{2}}{T_{1}}			&			0.986	\pm	0.006  	 		&	0.971	^{+	0.020	}_{-	0.019	}	&		\alpha (R_{\sun})			&	1.63	\pm	0.01	\\
R_{r}=\frac{R_{2}}{R_{1}}			&			0.466	\pm	0.006  	 		&	0.472	^{+	0.012	}_{-	0.012	}	&		T_{1} (K)		        		&	6696	\pm	202	\\
r_{1}	                    		&			0.567	\pm	0.009  	 		&	0.572	^{+	0.011	}_{-	0.012	}	&		T_{2} (K)		        		&	6502	\pm	234	\\
r_{2}	                    		&			0.264	\pm	0.008  	 		&	0.270	^{+	0.005	}_{-	0.006	}	&		M_{1} (M_{\sun})				&	1.19	\pm	0.02	\\
$\Omega_{12}$	            		&			2.065	\pm	0.008 	 		&	2.050	^{+	0.047	}_{-	0.041	}	&		M_{2} (M_{\sun})				&	0.19	\pm	0.02	\\
$i({\degr})$	            		&			88.5	\pm	1.2	     		&	85.7	^{+	2.6	}_{-	2.9	}	&		R_{1} (R_{\sun})				&	0.93	\pm	0.02	\\
f($\%$)	                    		&			63	    \pm	8  	    		&	72	^{+	14	}_{-	17	}	&		R_{2} (R_{\sun})				&	0.44	\pm	0.01	\\
\frac{L_{1B}}{L_{Btot}}	    		&			0.837	\pm	0.002	 		&	0.846	^{+	0.013	}_{-	0.013	}	&		L_{1} (L_{\sun})				&	1.45	\pm	0.07	\\
\frac{L_{1V}}{L_{Vtot}}	    		&			0.835	\pm	0.002	 		&	0.841	^{+	0.010	}_{-	0.011	}	&		L_{2} (L_{\sun})				&	0.31	\pm	0.05	\\
\frac{L_{1R}}{L_{Rtot}}	    		&			0.833	\pm	0.002  	 		&	0.839	^{+	0.009	}_{-	0.009	}	&		M_{{\rm bol}, 1} (mag)		        	&	4.34	\pm	0.05	\\
\frac{L_{1I}}{L_{Itot}}	    		&			0.831	\pm	0.002  			&	0.837	^{+	0.008	}_{-	0.008	}	&		M_{{\rm bol}, 2} (mag)		    		&	6.01	\pm	0.18	\\
\tableline
\multicolumn5c{ETV model}\\
\tableline
P_{3} ({\rm yr})	&	 \multicolumn2C{29^{+12}_{-9}}	& T_{p} (days) & 2457542\pm123 	\\
$A$ ({\rm days})	&	 \multicolumn2c{0.0058$\pm$0.0014}	& Q	 ({\rm days}~{\rm cycle}^{-1}) &	(-4.0$\pm$0.1)\times 10^{-11}\\	
a_{b} \sin i_{3} ({\rm AU})	&	 \multicolumn2c{1.2$\pm$0.3} & f(M_{3}) (M_{\sun})	& 0.00022\\
e_{3}	&	 \multicolumn2c{0.68$\pm$0.10}	& M_{3} (M_{\sun})	(90\degr)	&	0.18\\  	
$\omega_{b} ({\degr})$ & \multicolumn2c{328$\pm$10} & a_{3} (AU)(90\degr)	& 9.4\pm$2.3$ \\
\enddata
\tablecomments{$q$, $T_{r}$, $R_{r}$ are the mass, temperature, and radius ratios of the two components, respectively, $\frac{L_{1j}}{L_{jtot}}$ is the fractional luminosity of the primary component in filter $j$, $i$ is the orbital inclination, $\Omega_{12}$ is the potential of the components, $r_{1}$ and $r_{2}$ are the mean relative radii and $f=\frac{\Omega-\Omega_{\rm in}}{\Omega_{\rm out}-\Omega_{\rm in}}$ is the fillout factor, where  $\Omega_{\rm in}$ and $\Omega_{\rm out}$ are the modified Kopal potential of the inner and the outer Lagrangian points, respectively. A is the semiamplitude of LTTE, $\omega_{3}=\omega_{b}-\pi$ is the argument of the periastron of the third body and $a_{3}$ denotes the semimajor axis of the potential third body around the center of mass of the triple system, for $i_{3}=90\degr$} . The rest of the parameters of the ETV model are described in the text.
\end{deluxetable*}  

\section{Light Curve modeling} \label{sec:Analysis}
Our new $BVRI$ light curves of CRTS$\_$J163819 (Figure~\ref{fig:contours_qi_plots}c) show a typical W~UMa type EB system with total eclipses. Given that radial velocity curves are not available, we employed the \textsc{phoebe-0.31}a scripter \citep{2005ApJ...628..426P} to analyze our four-color LCs simultaneously and derive a reliable photometric mass-ratio, $q=\frac{M_{2}}{M_{1}}$  taking advantage of the totality \citep{2005Ap&SS.296..221T,2013CoSka..43...27H,2016PASA...33...43S}. Assuming circular orbits under ``Overcontact not in thermal contact'' mode, we performed a detailed scan in the mass ratio~- inclination ($q-i$) plane using a grid of predefined values. The mass ratio was selected in the range of [0.1-3.0] with a 0.01 resolution, and the inclination in the range of [$68\degr-90\degr$] using 1$\degr$ steps. The LCs were weighted according to their errors. 
The effective temperature of the primary (star eclipsed at phase zero) was set equal to the system's temperature, $T_{\rm eff}=6662\pm162$~K, as given by the Transiting Exoplanet Survey Satellite ({\em TESS }) Input Catalog Version 8 \citep[TIC-8;][]{2019AJ....158..138S}. The adopted gravity darkening coefficients and bolometric albedos were $A_{1,2}=0.5$ \citep{1973AcA....23...79R} and $g_{1,2}=0.32$ \citep{1967ZA.....65...89L}, respectively, and the bolometric and bandpass limb-darkening coefficients were interpolated from \cite{1993AJ....106.2096V} tables with a logarithmic law. In order to reach convergence, the Method of Multiple Subsets \citep{1976A&A....48..349W} was used, adjusting the passband luminosity $L_{1}$ and the temperature of the secondary component ($T_{2}$) or the modified surface equipotentials ($\Omega_{12}=\Omega_{1}=\Omega_{2}$) for 20 iterations each set. Finally, parameters $\Omega_{12}$, $T_{2}$ and $L_{1}$ were adjusted together to converge for 50 iterations. 

We follow the fitting strategy of \citet{2022MNRAS.512.1244C}. 
The corresponding photometric parameters from our light curve modeling, along with their errors calculated via a Monte Carlo (MC) procedure, are listed in column 2 of Table~\ref{tab:Tab1}. The mass ratio ($\log q$) versus cost function value ($\log \chi^2$) derived from the $q-i$ scan method is shown in  Figure~\ref{fig:contours_qi_plots}a.

To explore further the parameter space and the uncertainties of the model parameters, we used a genetic algorithm optimizer technique. Specifically, we apply \textsc{pikaia} genetic algorithm \citep[GA,][]{1995ApJS..101..309C} interfaced with {\textsc{phoebe}} as adapted in \cite{PapageorgiouPhD} and \cite{2023AJ....165...80P}. The population size was set to 120 individuals, and 2000 generations were computed. The best-fit parameters as defined by the best individuals from this list are $q=0.16\pm0.012$, $i=85.3{\degr}\pm3.5{\degr}$, $T_2/T_1=0.972\pm0.026$, $\Omega_{12}=2.062\pm0.028$, $r_{1}=0.570\pm0.009$, and $r_{2}=0.271\pm0.017$. Furthermore, to determine the uncertainties more robustly, we used the affine invariant Markov Chain Monte Carlo (MCMC) ensemble sampler implemented in the \textsc{ EMCEE} \citep{2013PASP..125..306F} Python package, coupled with the 2015 version of the Wilson-Devinney (W-D) binary star modeling code \citep{WD1971, 2020ascl.soft04004W}.$\footnote{\url{ftp://ftp.astro.ufl.edu/pub/wilson/lcdc2015/}}$ The values of $q, i, T_{r}$, $\Omega_{1,2}$, and $L_{1}$, as determined by $q-i$ scan and genetic algorithm optimizer, were served as priors for the MCMC sampling. A total of 30 walkers were used. The MCMC parameter search was run for $6,\!000-10,\!000$ steps with a burn-in phase of $1,\!500-2,\!000$ steps, for our $BVRI$ LCs, resulting in $\sim 180,\!000$ iterations. Figures~\ref{fig:contours_qi_plots}b and 1c show the probability distributions of $q, i, T_{r}$, $\Omega_{1,2}$ and the theoretical synthetic models (solid lines), respectively. The final parameters with their uncertainties listed in the third column of Table~\ref{tab:Tab1} are obtained using the mean value and the standard deviation.    

\section{Period Variation Analysis} \label{sec:PeriodAnalysis}
\begin{figure*}[!htb]
\gridline{\fig{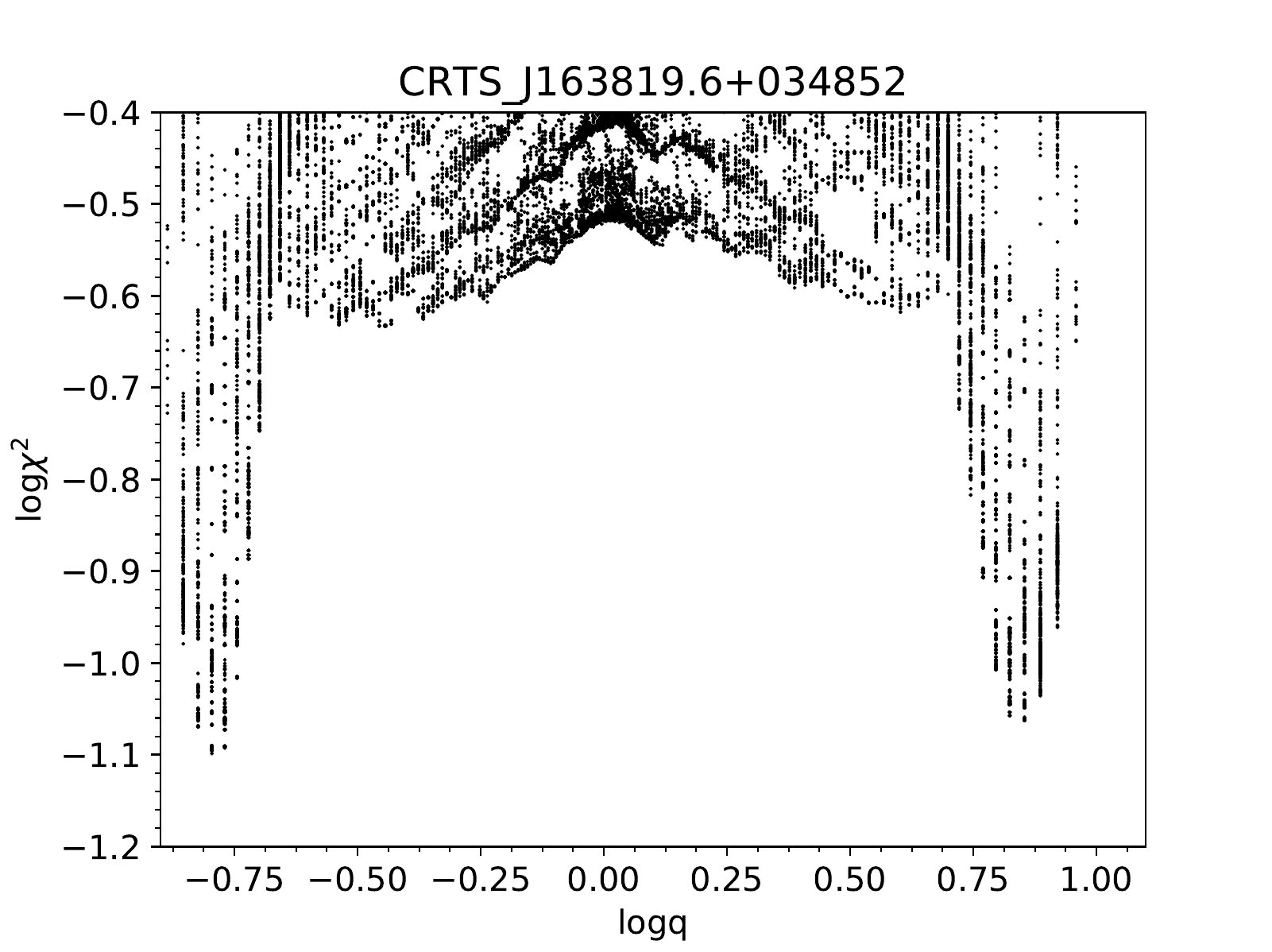}{0.5\textwidth}{(a)}
          \fig{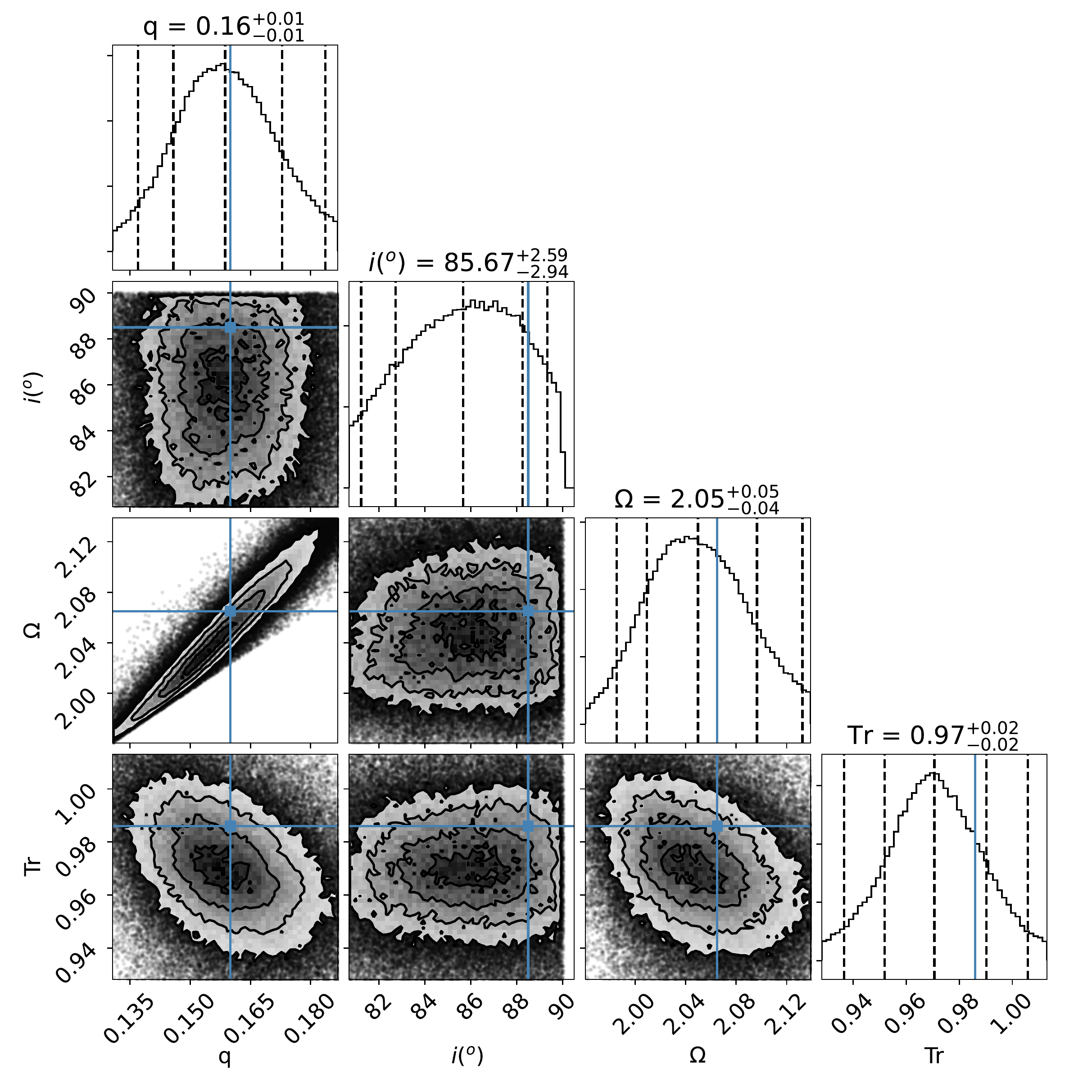}{0.5\textwidth}{(b)}}
\gridline{\fig{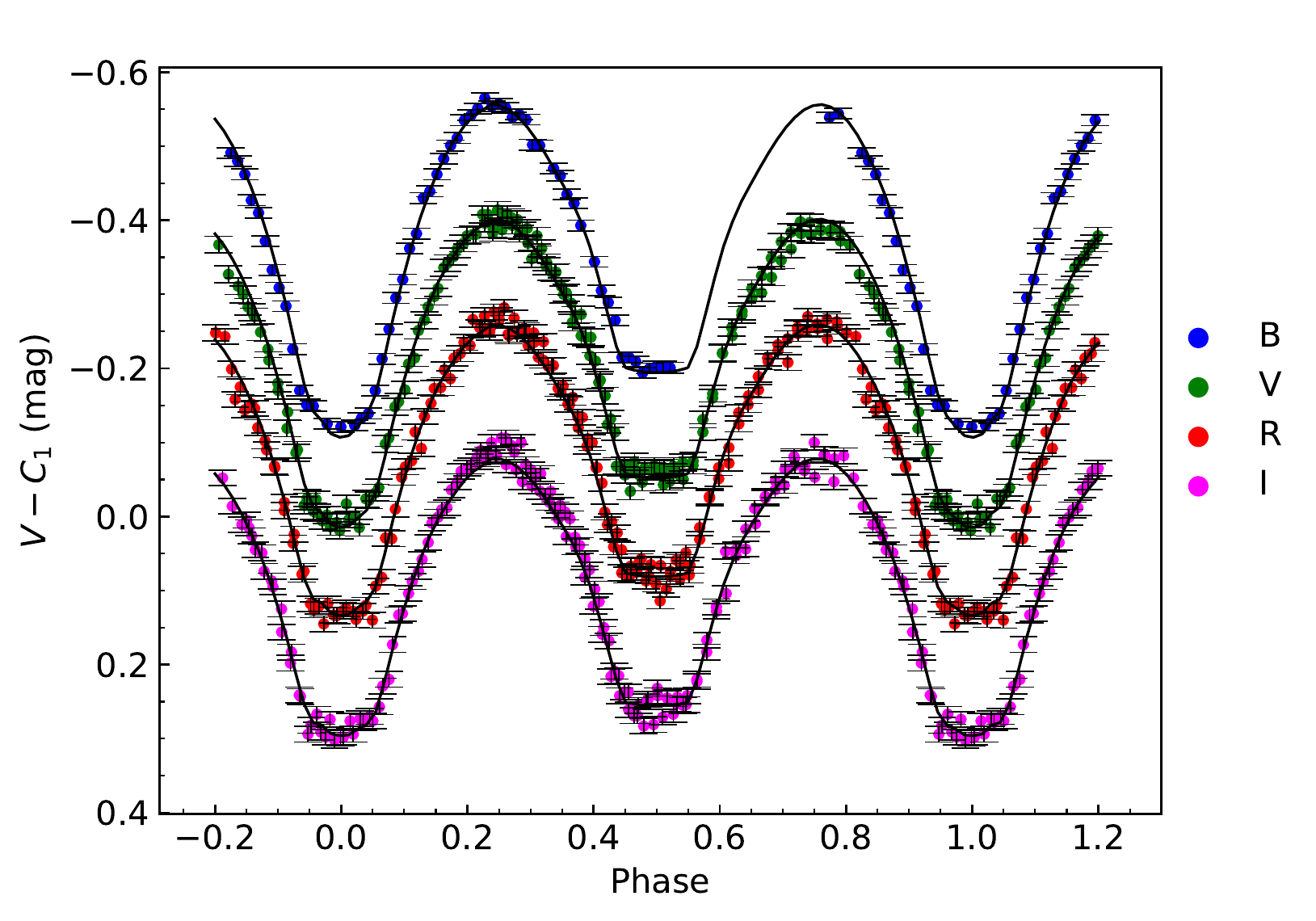}{0.5\textwidth}{(c)} 
           \fig{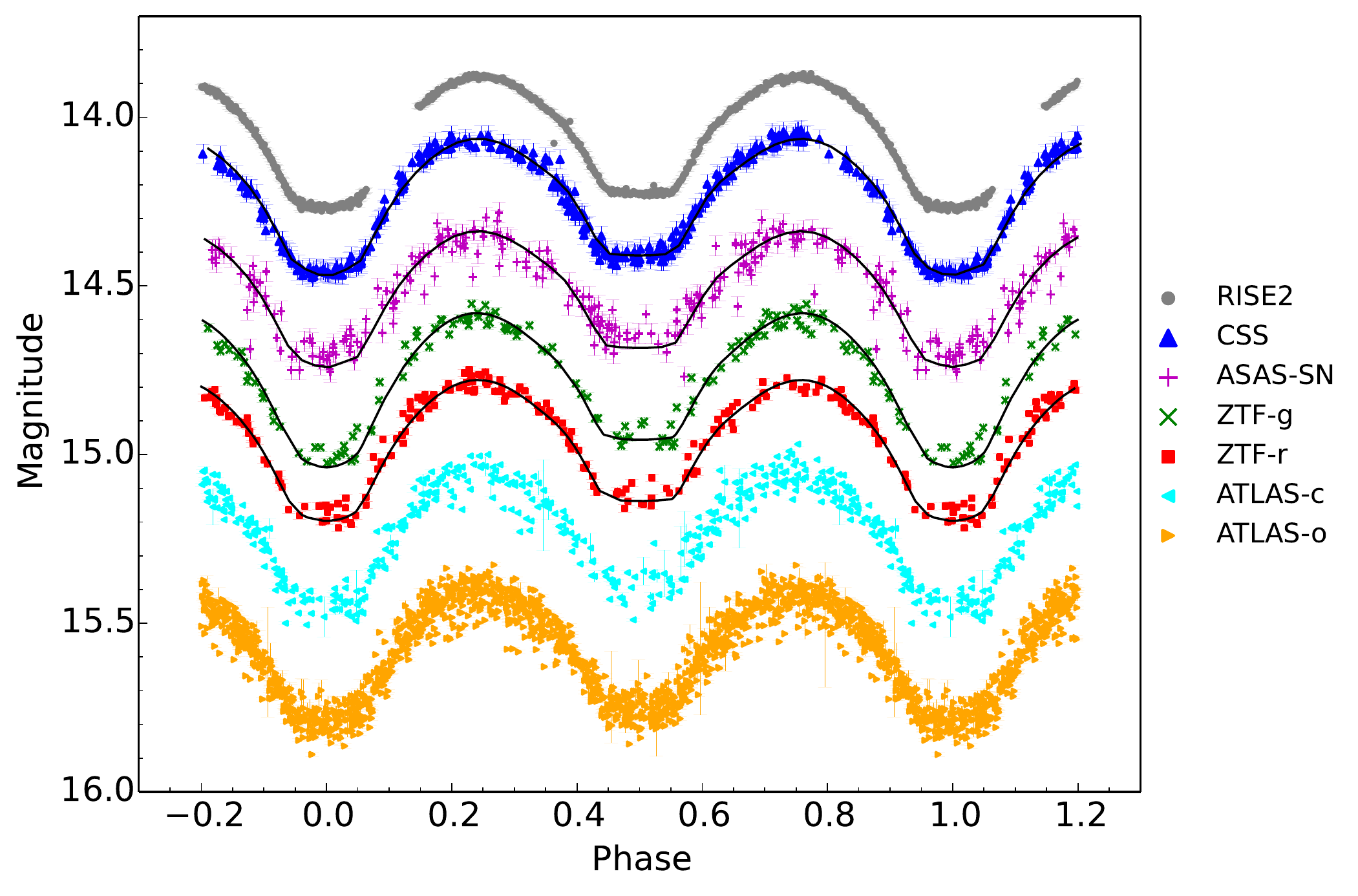}{0.50\textwidth}{(d)}
}
\caption{a)~Mass-ratio ($\log q$) versus cost function value ($\log \chi^2$) derived from the $q-i$ scan method, as applied to CRTS$\_$J163819. b)~The probability distributions \citep{2013PASP..125..306F} of $q, i, \Omega$, and $T_{r}$, determined by the MCMC modeling. The blue lines represent the solution of the system from the $q-i$ method. c)~The new observed folded light curves from Aristarchos telescope and synthetic models of CRTS$\_$J163819 (solid lines). 
The $B,I$ light curves are shifted vertically for clarity, by $B+0.2$, and $I+0.1$ respectively. d)~Photometric data (with errors) of CRTS$\_$J163819.6, folded with respect to the period of 0.205332~d.} 
\label{fig:contours_qi_plots}
\end{figure*}

Due to the absence of available times of minimum light (ToM) of CRTS$\_$J163819, we exploited the information provided from large astronomical surveys (CSS, ASAS-SN, ZTF, ATLAS) by collecting the publicly available LCs. This information not only provides insight on the LC evolution history but also valuable information on the period variation in a time span of more than 15~yr (2004-2021). Figure~\ref{fig:contours_qi_plots}d shows the archival LCs of CRTS$\_$J163819 in different passbands, mined from CSS, ASAS-SN, ZTF, and ATLAS, and the LC from the broad  RISE-2 VR-filter observed on July 3 2018. The model derived in Section~\ref{sec:Analysis} is overplotted on CSS and ASAS-SN in V band. The models for ZTF $g$ and $r$ bands were constructed using \textsc{phoebe-2} \citep{2016ApJS..227...29P} with our solution parameters. The RISE-2 and ATLAS LCs were used only to derive 88 ToM, since they are obtained in VR, ``cyan'' (c) and ``orange'' (o) wide filters respectively.  

\begin{figure*}[!htb]
\plotone{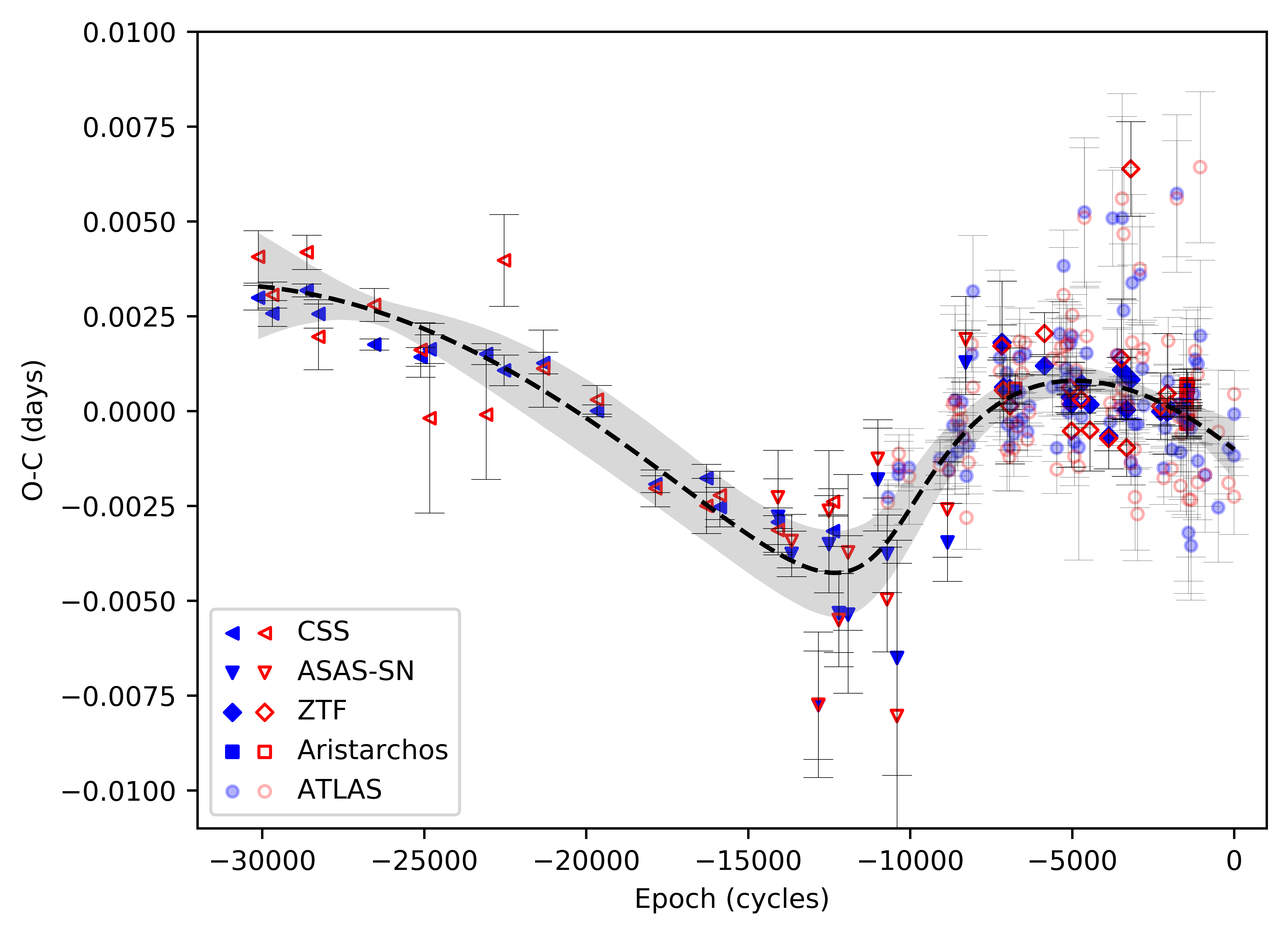}
\caption{$O-C$ diagram of CRTS$\_$J163819, computed with respect to the linear terms of equation~\ref{eq:OC1}. The dashed line represents the full contribution of the quadratic plus LTTE ephemeris, with the shading being indicative of the 1$\sigma$ uncertainty. Blue filled and red hollow symbols represent the primary and secondary ToM, respectively.\label{fig:OCplots} 
}
\end{figure*}

Instead of using phenomenological models for ToM calculations that have been proven efficient when dealing with a large number of LCs \citep{2018MNRAS.480.4557L,2019MNRAS.485.2562H,2021MNRAS.503.2979P}, we take advantage of the constructed models, for each archival and new photometric dataset and we calculate the ToM by applying the semi-automatic fitting procedure \citep[AFP;][]{2014A&A...572A..71Z}. All 275 ToM are converted to Heliocentric Julian Dates (HJD). In the case of surveys with low time resolution (low cadence), we used longer time span data to construct the full phase LC. Therefore, a linear ephemeris for CRTS$\_$J163819 was derived using the new ToM, as follows:  
\begin{equation}
    {\rm MinI} = 2459692.76790 + 0.2053321 \times E, \label{eq:OC1}
\end{equation}
where MinI is the HJD time of minimum at epoch $E$. To study the orbital period variation for the first time we calculate all available ToM with  equation~\ref{eq:OC1} and estimate the observed minus computed ($O-C$) times of minima  (Figure~\ref{fig:OCplots}). As can be seen in this figure, the $O-C$ curve shows a
combination of a downward parabola and a nearly sinusoidal variation. Therefore, we incorporate in the eclipse times variation (ETV) model both a parabolic and a light-travel time effect (LTTE) term following \citet{Irwin1952}, due to mass transfer and a possible tertiary companion, respectively: 
\begin{equation}
O-C=\Delta{T_{\rm o}}+\Delta{P_{\rm o}}\times E + Q \times E^{2}+ \tau_{3}, 
    \label{eq:OC2}
\end{equation}
\noindent where $\Delta{P_{\rm o}}$ and $\Delta{T_{\rm o}}$ are the corrections of the initial period and primary minimum with respect to the values in equation~\ref{eq:OC1}, $Q$ is the long-term rate of change in orbital period, and  $\tau_{3}$ is the LTTE due to a circumbinary companion \citep{Irwin1952}. The latter includes the projected semimajor axis $\alpha_{b}sini_{3}$ (or semi-amplitude of the LTTE, $A$), eccentricity $e_{b}$, argument of the periastron of the orbit $\omega_{b}$, true anomaly $\upsilon_{b}$ of the position of the mass center, and time of periastron passage $T_{p}$, all of which referring to the  EB's center of mass around the center of mass of the triple system. 
The period $P_{3}$ and the time of periastron passage $T_{p}$ are included when solving Kepler's equations.

To solve for the eight parameters ($\Delta{T_{o}}$, $\Delta{P_{o}}$, $Q$, $A$, $e_{b}$, $\omega_{b}$, $P_{3}$, $T_{p}$) we initialize the model parameters by fitting the ETV model coupled with the \textsc{pikaia} genetic algorithm. We evolved a population of 120 sets of parameters randomly generated from uniform distributions, for 1,000 generations. The mean parameter values from the last generation were used as input parameters, and a least-squares (LS) fitting was performed. The model rapidly converged to a solution within 1$\sigma$ of the GA initial solution. We estimate the model parameter errors by performing a final fitting via an MCMC procedure. This was done by using the \textsc{ pyMC} \citep{2015ascl.soft06005F} package in Python. The model parameters were sampled from uniform distributions centered at the LS solution, with $2\sigma$ ranges as provided by the GA solution.  To avoid biases in the initial solution, a burn-in period was set to 20,000 iterations (200,000 iterations in total). Figure~\ref{fig:OCplots} shows the final model, overplotted on the $O-C$ diagram. The derived parameters are listed in  Table~\ref{tab:Tab1}. 


\epsscale{1.3}
\begin{figure*}[ht!]
 
\gridline{\fig{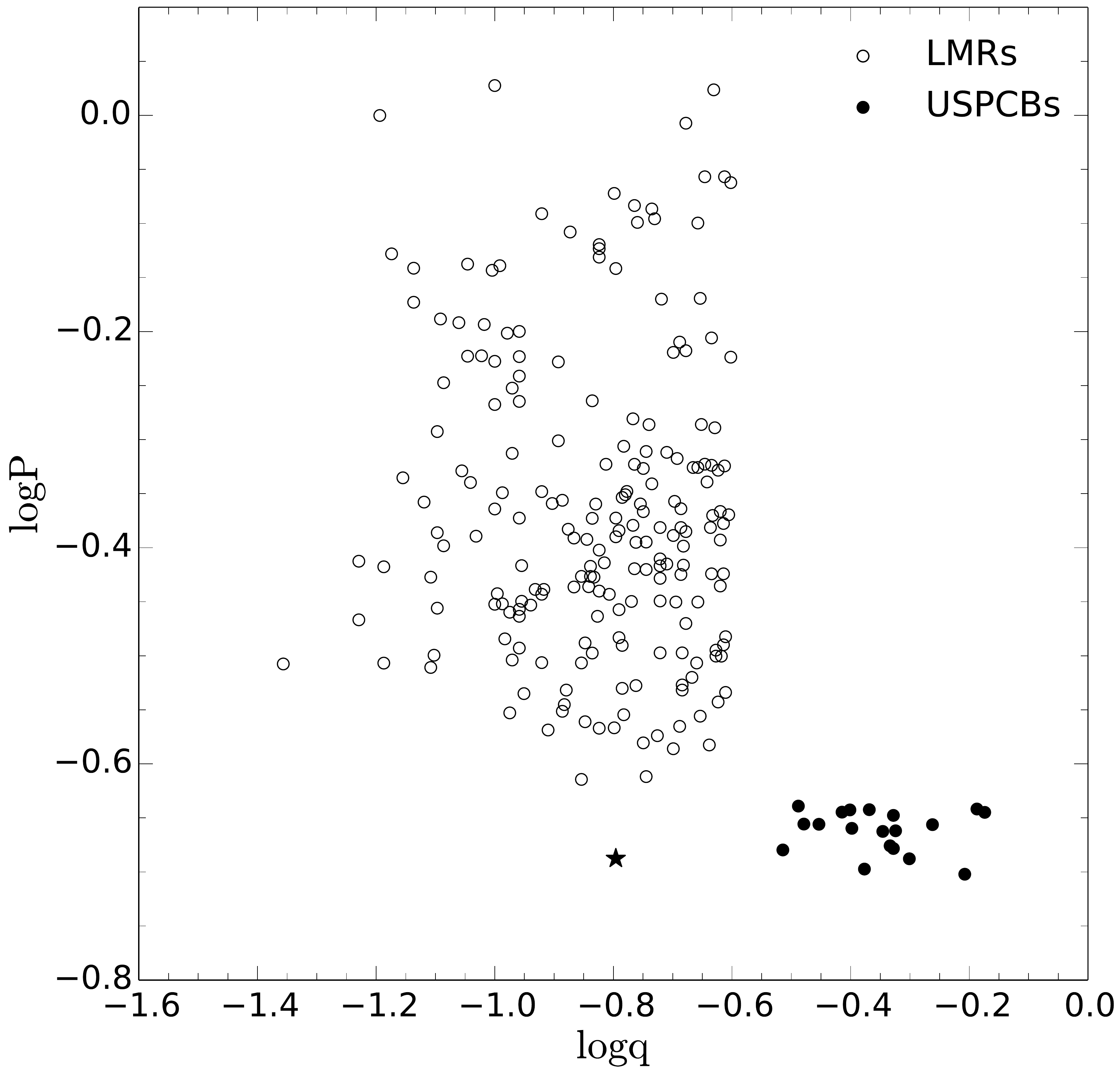}{0.5\textwidth}{(a)}
          \hspace{0.5cm}
          \fig{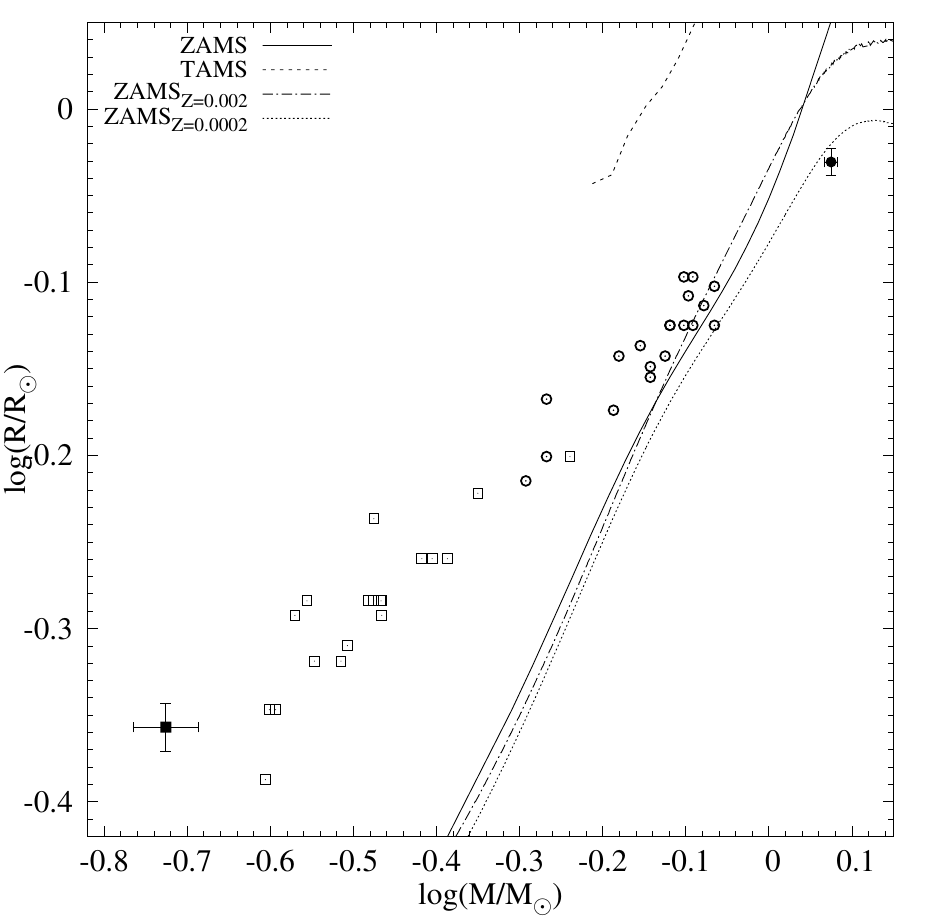}{0.5\textwidth}{(b)}}
\caption{a)~Distribution of LMRs \citep[open circles, from][]{2022MNRAS.512.1244C} and USPCBs  \citep[filled circles, from][]{2023AJ....165...80P} in the  $\log q-\log P$ plane. The star symbol represents CRTS$\_$J163819. ~b) The primary and secondary components (filled circles and squares, respectively) of CRTS$\_$J163819 compared with primary and secondary components (open circles and squares, respectively) of USPCBs \citep{2023AJ....165...80P}, plotted on the $\log {M}-\log {R}$ diagram. ZAMS loci for different metallicities (solid, dash-dotted, and dotted lines), along with a TAMS locus (dashed line) for solar metallicity, as obtained using the BSE code \citep{Hurley2002}, are overplotted.}\label{fig:evolution}
\end{figure*}

\section{Results} \label{sec:Results}
  The photometric parameters determined from different methods ($q-i$ search, heuristic scan with parameter perturbation/MC with \textsc{phoebe}, MCMC sampler with W-D, and \textsc{pikaia} genetic algorithms with \textsc{phoebe}) are consistent,  within the reported errors.  To determine the masses $(M_{1},M_{2})$, radii $(R_{1},R_{2})$, and luminosities $(L_{1},L_{2})$ of the components, we use the results and the uncertainties of the MCMC sampler following the method proposed in \cite{2022MNRAS.512.1244C} using the distance given by {\em Gaia} Data Release 3 \citep[DR3;][]{GaiaDR3} as $939 \pm 15$~pc. 
  The primary's mass was estimated by the mass-luminosity approximation \citep[for details see; ][]{ 2022MNRAS.512.1244C,2023AJ....165...80P}:
  \begin{equation}
\label{eq:1}
\log L_{1}=\log (0.63 \pm 0.04)+(4.8 \pm 0.2)\log M_{1} . 
\end{equation}
\noindent
The approximate absolute parameters of the binary components are listed in the fifth column of Table~\ref{tab:Tab1}.

If the periodic component of CRTS$\_$J163819 eclipse timings is interpreted as the  LTTE due to a circumbinary companion, 
the minimum mass ($i_{3}=90{\degr}$) is found to be $M_{3}=0.18 \, M_{\sun}$
at a separation from the binary $\sim$11.0~AU ($\alpha_{b}+ \alpha_{3}$ for $ i_{3}=90{\degr}$ Table~\ref{tab:Tab1}). If it is a main-sequence star, it would have a temperature of 3100~K  and a bolometric luminosity $0.003 \, L_{\sun}$. Thus, the contribution to the total light of the system would be $L_{3}/(L_{1}+L_{2}+L_{3})=0.002$. Accordingly, in agreement with our results, the tertiary companion may not be detected as a third light source in the LC analysis. We computed also the semiamplitude of the third body dynamic perturbation on the binary orbit, finding it to be negligible. 
The quadratic term in the analysis of the $O-C$ diagram represents a continuous decrease at a rate of $-7.11\times 10^{-8} \, {\rm days}~{\rm yr}^{-1}$. Such a variation could be the result of mass transfer from the more massive component (primary) to its  secondary companion, or angular momentum loss due to a magnetic stellar wind. Considering conservative mass transfer, the transfer rate  is $dM_{1}/dt=-2.61\times 10^{-8} M_{\sun}~{\rm yr}^{-1}$, typical for mass-transferring USPCBs \citep{2020AJ....159..189L} and LMRs \citep{2021ApJ...922..122L}. Assuming that the transfer is produced on a thermal time scale, $\tau_{\rm th}=3.24\times 10^{7} \, {\rm yr}$ , the mass is transferred  to the companion at a rate of $M_{1}/\tau_{\rm th}=3.67\times10^{-8} M_{\sun}~{\rm yr}^{-1}$. This value is similar to that calculated from the period analysis, making mass transfer a possible cause. We also examined angular momentum loss (AML) via a magnetized wind as a possible mechanism for the period decrease, using equation~(23) of \cite{1995MNRAS.274.1019S}. We found the AML rate to be two orders of magnitude smaller than the observed $dP/dt$ value. This suggests that magnetic braking is not the main cause of period decrease.

An alternative explanation of the orbital period modulation of CRTS$\_$J163819 system is strong magnetic activity due to the Applegate effect \citep{1992ApJ...385..621A}.
Using the \citet{2016A&A...587A..34V} eclipsing time variation calculator\footnote{\url{http://theory-starformation-group.cl/applegate/index.php}} and the absolute physical parameters of the components given in Table~\ref{tab:Tab1}, we found that the energy $\Delta E$ required to drive the Applegate mechanism is 29\% of the available energy $E_{\rm sec}$, produced in the magnetically active secondary star for the finite-shell two-zone model  \citep{2016A&A...587A..34V}, or 1.5\% for the thin-shell model \citep{2009Ap&SS.319..119T}. Therefore, for  CRTS$\_$J163819, since $\Delta E/E_{\rm sec}<1$, the Applegate mechanism appears to be  energetically feasible. However, for a rapidly rotating secondary with a convective envelope, we expect signs of dynamo activity, such as starspots (or other chromospheric manifestations), which typically manifest themselves in the LCs by an asymmetry between maxima (O'Connell effect). We did not detect the presence of unequal maxima in the LCs or any source of UV or X-ray emission within 2\arcsec\ of the position of CRTS$\_$J163819 in the relevant catalogs.

However, we have to note that since the time span of the available data from the astronomical surveys covers the half cycle of the total outer orbit, we consider this model as highly preliminary to the outer orbit determination. 

\section{Discussion and Conclusions} \label{sec:Discussion}
The first photometric solution revealed that CRTS$\_$J163819  has an extremely low mass ratio, $q=0.16 \pm 0.01$, which makes this system unique among LMRs and USPCBs, given that its period (0.205332~d) is also under the contact period limit of 0.22~d (Figure~\ref{fig:evolution}a). The fillout factor ($72\pm 15 \%$) suggests that the system is in deep contact, and the temperature difference between the components indicates that they are under thermal contact. It appears to be an A subtype EB, in which a primary of $F5/F4$ spectral type is eclipsed during the deeper minimum. We note, however, that spectroscopic studies of this system will ultimately be needed to more robustly determine the mass ratio more. In this regard, we note that there is generally very good agreement between photometric and spectroscopic mass ratios in totally eclipsing contact EBs \citep{2021ApJ...922..122L}. Though, we should point out that according to \cite{2020AJ....160..104R},
the discrepancy between spectroscopic and photometric mass ratio can be as $10 \% $ and in some cases even larger (AW UMa). In other cases e.g $\epsilon$ CrA although the photometric mass ratio $q_{ph}=0.128\pm0.0014$ is within $4\sigma$ of the spectroscopic value $q_{sp}=0.13\pm 0.001$ and the discrepancy is minor, the detection of similar complex velocity flows as in AW UMa, may imply the necessity to modify the Roche lobe-based photometric mode \citep{2020AJ....160..104R}.
 
Figure~\ref{fig:evolution}b shows the position of the two components of  CRTS$\_$J163819 on the $\log M-\log R$ diagram, along with other USPCBs compiled by \cite{2023AJ....165...80P}. The zero-age main sequence (ZAMS) and the terminal-age main sequence (TAMS), calculated for different metallicities using the Binary Star Evolution code \citep[BSE,][]{Hurley2002}, are overplotted. According to the locations of USPCBs primaries with respect to the ZAMS, the more massive primary star of CRTS$\_$J163819 seems to act as a normal main-sequence star with relatively low metallicity.  The secondary, less massive component is located beyond the TAMS, in accordance with the positions of other USPCB secondaries.

We also evaluated the potential progenitor of CRTS$\_$J163819 to be an ordinary detached system using the evolutionary model of \cite{2006Stepien} (details of the models' construction are provided in Appendix \ref{appendix}). 

Although the structure and detailed evolutionary process during the contact phase are still open questions, we can only point out that CRTS$\_$J163819, having very low $q$ and $P$, must be at a late stage of evolution. As it has a high fillout factor ($\geq 72\%$), we can consider it as the progenitor of a merger that will ultimately lead to an FK~Com-like fast-rotating single star, blue straggler, or red nova \citep[see, e.g.,][]{1995ApJ...444L..41R,2015A&A...577A.117S,2011A&A...528A.114T}. 
To investigate its stability, we calculate the ratio of the spin angular momentum ($J_{\rm s}$) to the orbital angular momentum ($J_{\rm o}$) (assuming corotation of the components spin and the orbit) to be $J_{\rm s}/J_{\rm o}= 0.145$, using values of $k_{1}^2= k_{2}^2=k^2=0.06$ \citep{1995ApJ...444L..41R} for the gyration radii of both components. Alternatively,  $J_{\rm s}/J_{\rm o}=0.154$ using $k_{2}^2\sim 0.205$ and evaluating $k_{1}^2=0.058$ from \cite{2022MNRAS.512.1244C} for a star of $1.19 \,M_{\sun}$. In both cases, $J_{\rm s}/J_{\rm o}$ does not exceed the instability value of $\sim 1/3$ \citep[Darwin's instability; see, e.g.,][and references therein]{Hut1980}, 
indicating that the system is in a stable state at present. Nevertheless,  CRTS$\_$J163819 has the highest value of $J_{\rm s}/J_{\rm o}$ amongst USPCBs (see Appendix \ref{appendix}). To further investigate its stability, we calculate the theoretical instability mass ratio ($q_{\rm inst}=0.088$), the instability separation ($a_{\rm inst}=1.269 \, R_{\sun}$) and the instability period ($P_{\rm inst}=0.141$~d) by applying the method of \cite{2021MNRAS.501..229W}, using $k_{2}^2\sim 0.205$ and $k_{1}^2=0.058$ as above. All three resulting parameters are smaller than the corresponding current parameters, suggesting that CRTS$\_$J163819 is a stable binary for now.

 According to this approach, there is no single value to the limit of the mass instability ratio, as it depends on the primary mass for low-mass stars, and therefore on the different structures of tidally deformed and rotating ZAMS stars, metallicity, and age. However, the inability to explain why systems such as V1187~Her and V857~Her violate the theoretical limits highlights the need to extract more reliable mass ratios from radial velocity curves and/or the need to understand and model the stationary flow structures \citep{2020AJ....160..104R}. 
 Why is CRTS$\_$J163819 the only such system to be observed so far? Why are systems with this combination of $q$ and $P$ so rare? To investigate these issues, we need future spectroscopic observations of the radial velocities of the components to establish the mass ratio. Furthermore, the measurement of the radial velocity would be useful to compute the space motions of CRTS$\_$J163819 to infer its kinematical age through age-velocity dispersion relations.




\begin{acknowledgments}
Based on observations made with the 2.3m Aristarchos telescope, Helmos Observatory, Greece, which is operated by the Institute for Astronomy, Astrophysics, Space Applications and Remote Sensing of the National Observatory of Athens, Greece. This research is co-financed by Greece and the European Union (European Social Fund- ESF) through the Operational Programme ``Human Resources Development, Education and Lifelong Learning'' in the context of the project ``Reinforcement of Postdoctoral Researchers - $2^{nd}$ Cycle'' (MIS-5033021), implemented by the State Scholarships Foundation (IKY). EL gratefully acknowledges the support provided by IKY ``Scholarship Programme for PhD candidates in the Greek Universities''.
MC and CEFL acknowledge the support provided by ANID's Millennium Science Initiative through grant ICN12\textunderscore 12009, awarded to the Millennium Institute of Astrophysics (MAS). Additional support for MC is provided by ANID Basal project FB210003.
This research has made use of the VizieR catalogue access tool, CDS, Strasbourg, France.\\
We would like to thank Prof. K. Stepien for his useful comments/suggestions on the models of progenitors.\\
We would like to thank the referee for constructive comments and recommendations that have improved the paper.\\


\end{acknowledgments}

%

\vspace{5mm}
\facilities{Aristarchos 2.3 m, ZTF, ASSAS-SN, CSS}


\software{\textsc{PHOEBE-0.31a} \citep{2005ApJ...628..426P} , 
          \textsc{PHOEBE-2} \citep{2016ApJS..227...29P}
          \textsc{EMCEE} \citep{2013PASP..125..306F},
          \textsc{W-D code} \citep{2020ascl.soft04004W},
          \textsc{triangle.py-v0.1.1}\citep{dan_foreman_mackey_2014_11020}
          \textsc{pyMC} \citep{2015ascl.soft06005F}
          \textsc{MWDUST} \citep{2016ApJ...818..130B},
          \textsc{PyRAF} \citep{2012ascl.soft07011S},
          \textsc{Astrometry.net} \citep{2010AJ....139.1782L}
          }



\appendix
\section{Models of progenitors of CRTS$\_$J163819}\label{appendix}

\begin{figure*}[ht!]
\gridline{\fig{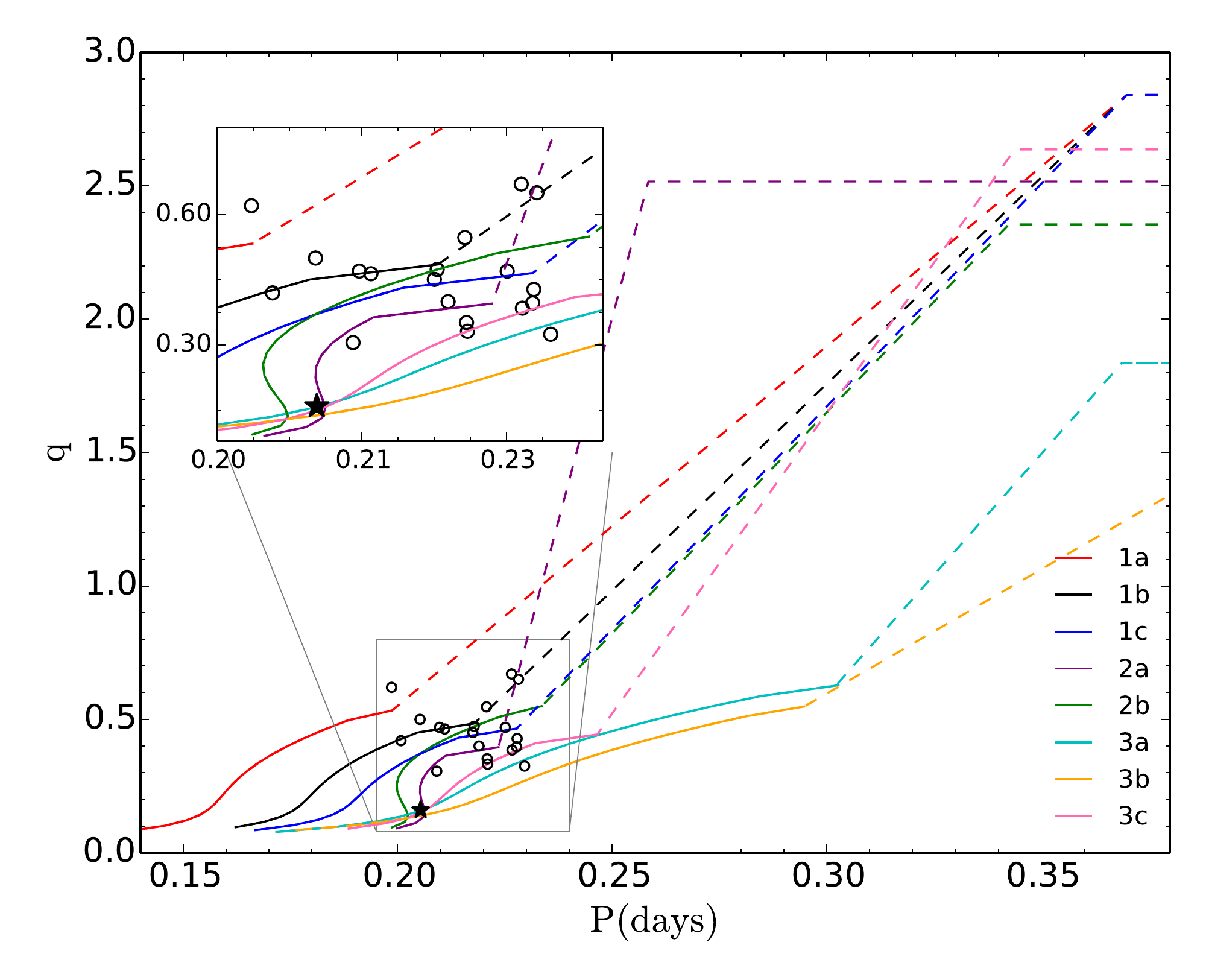}{0.5\textwidth}{(a)}
          \hspace{0.5cm}
          \fig{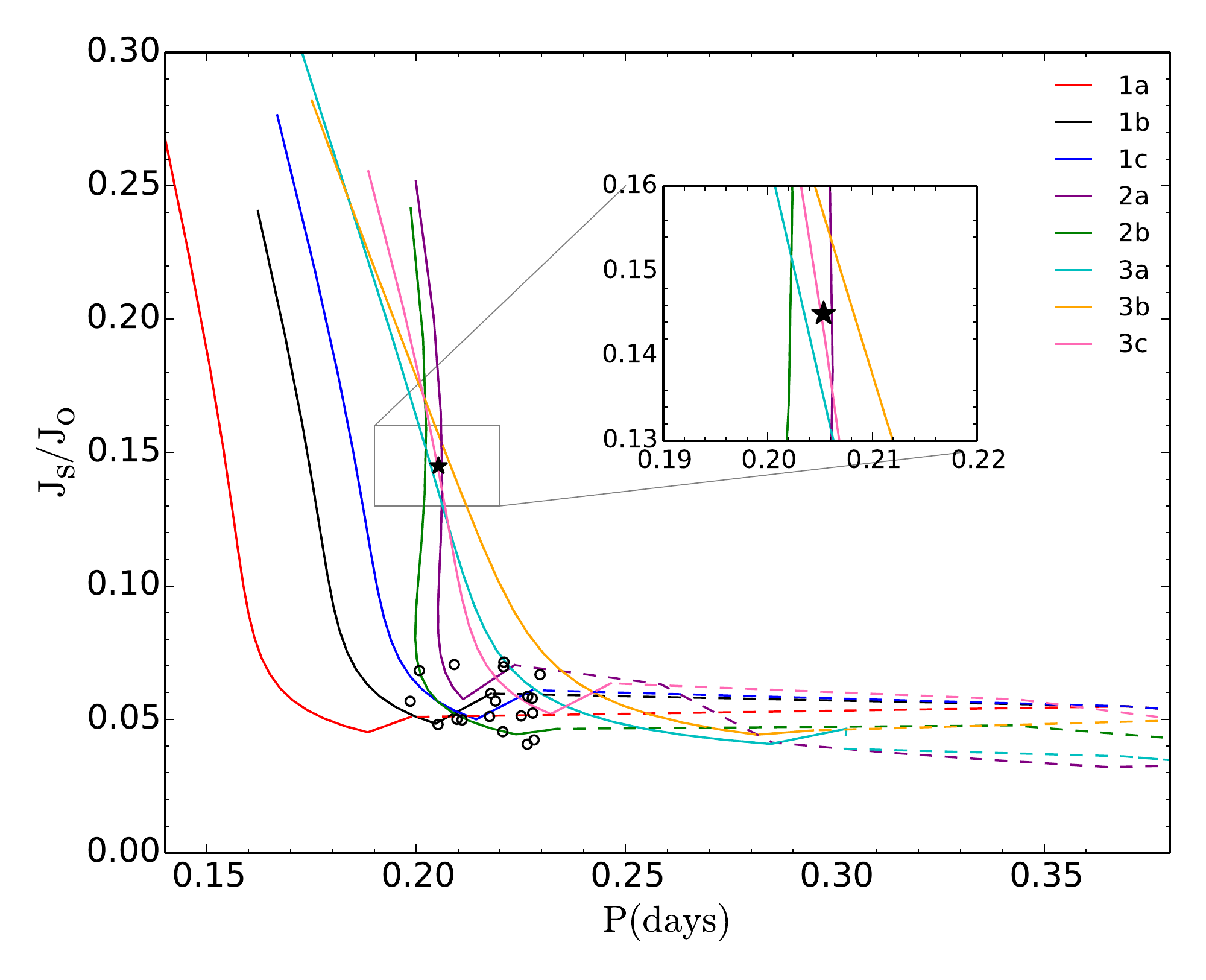}{0.5\textwidth}{(b)}}
\caption{(a) The mass ratios q and (b) the ratios of spin to orbital angular momentum $J_{\rm s}/J_{\rm o}$ of CRTS$\_$J163819 (filled star) and USPCBs  \citep[open circles, from][]{2023AJ....165...80P} are plotted as a function of the period $P$ (days), and compared with the various evolutionary models (colored lines, labeled as 1a-c, 2a-b, 3a-c, following the insets) as described in the Appendix\ref{appendix}. Contact phases correspond to the solid lines and pre-contact phases to the dashed lines.}\label{fig:Stepien}
\end{figure*}

Although there have been a few studies in the structure or evolution of contact binaries via the detached channel \citep{2005ApJ...629.1055Y, 2014MNRAS.438..859J, 2020MNRAS.492.2731J} taking into account the complex physical processes (spin, orbital rotation, tides, mass and energy exchange, angular momentum loss due to magnetic braking), for most of the models the examples were stopped as soon as contact was reached. To approximate the progenitors of the current system we adopted the scenario proposed by \cite{2006AcA....56..347S} and \citet[][and references therein]{2015A&A...577A.117S} that can predict the observational phenomena and the parameters of cool low mass contact binaries.
This scenario, describes the evolution of a detached binary of two magnetically active MS stars with circular, coplanar, and synchronized orbits from the zero-age main sequence (ZAMS) to a phase just before the components merge or form a common envelope. Neglecting any interaction between the winds or other mechanisms (possible tertiary companion), the dominating mechanisms of the orbit evolution are the magnetic braking due to the winds and the mass transfer between the components. Three phases make up the evolutionary calculations: the first phase considers a  detached binary, during which the more massive component (donor) evolves from ZAMS to the Roche lobe overflow (RLOF); the second phase considers a rapid, conservative mass exchange from donor to less massive component (accretor) and the third phase considers a slow mass transfer to the accretor as a result of nuclear evolution of the donor.The mathematical description of the model is given more fully in \citep[see, Eq. 1-8, section 2.1,][]{2015A&A...577A.117S}.  The set of the 8 basic equations of the model are the third Kepler law, the standard expression for binary rotational and orbital angular momentum, the following formulas that predict how much mass and angular momentum (AM) the winds carry away, and the approximate expressions for inner Roche-lobes sizes \citep{1983ApJ...268..368E}. 
\begin{equation}
\label{eq:A1}
\frac{dM_{\rm 1,2}}{dt}~[M_{\sun}~{\rm yr}^{-1}]=10^{-11}R_{1,2}^2  
\end{equation}
\begin{equation}
\label{eq:A2}
\frac{dJ_{\rm tot}}{dt}~[g~ cm^2 s^{-1}{\rm yr}^{-1}] =-4.9 \times 10^{41} (R_{1}^2 M_{1}+R_{2}^2 M_{2})/P,  
\end{equation}
where $M_{1,2}, R_{1,2}$ are the masses and radii of both components, and $J_{\rm tot}$ is the total (orbital and rotational) angular momentum. 
This scenario entails the following basic assumptions:
\begin{itemize}
\item because we lack detailed evolutionary models of binary components during and after the mass exchange, regardless of the details of the mass transfer process, both stars are assumed to retain thermal equilibrium at every step and phase. The set of 8 equations is integrated at every time step and all stellar parameters are interpolated from the PAdova and TRieste Stellar Evolution Code \citep[PARSEC;][]{2012MNRAS.427..127B} grid, thus from a single star
evolutionary model
\item equations~\ref{eq:A1},~\ref{eq:A2} are calibrated using observational information about the rotation of single, magnetically active stars of various ages and empirically calculated mass-loss rates of single, solar-type stars \citep{2006AcA....56..347S}. The constant in equation~\ref{eq:A1} is uncertain within a factor of 2 and that in equation~\ref{eq:A2} is uncertain to $\pm 30 \%$. The binary system can reach contact while both components are still on the MS (the massive component reaches RLOF just reaching TAMS) after a rapid mass exchange till the mass ratio reversal. 
\item the evolution path to contact is determined by a) the initial values of the masses of the progenitor cool detached binary and its initial period b) the adopted mass transfer rate at the first overflow (when the initial massive component of the detached phase fills the Roche lobe for the first time, RLOF) and after mass ratio reversal and c) the metallicity as it determines the MS lifetime and thus the time needed for the massive component to reach RLOF.  
\item  The first (rapid) constant mass transfer rate during Phase II (after RLOF of the initially more massive component)  is adopted based on the observation that the total mass transferred in any of the modeled binary did not exceed half solar mass, assuming that the mass is transferred at a thermal time scale of $10^{8}$ yrs. As \cite{2006AcA....56..347S} suggests after trying many formulas of mass transfer the best-adopted values that secured stability for most of the models were around $5 \times 10^{-9}\, M_{\sun}~{\rm yr}^{-1}$.
\item In phase III (beginning of contact configuration) the influence of AML balances the mass transfer rate so that the orbit remains tight. Thus the mass transfer rate resulted from the comparison of the radius of the accretor (secondary component)  to the Roche lobe in order to keep its contact configuration and the radius of the primary was assumed to increase a little due to evolutionary effects. The correct value of mass transfer rate required very fine tuning not only to maintain the contact configuration but also to shorten the orbital period. The mass is transferred at a rate proportional to the excess of the donor's size above the Roche lobe. The resulting values lie in the interval $3-4 \times 10^{-10}\, M_{\sun}~{\rm yr}^{-1}$ i.e. they are about ten times lower than in Phase II.
\end{itemize}
Although without an accurate binary age, its progenitor cannot be uniquely determined, the current observed total mass of CRTS$\_$J163819 restricts the initial total progenitor mass to the range of 1.35-1.4 $M_{\sun}$, because the expected mass loss due to winds is typically less than  $0.1\, M_{\sun}$ \citep{2015A&A...577A.117S}. In Figure~\ref{fig:Stepien}a, we present period tracks of old sets of models (1a-c and 2a-b) in the $P-q$ plane, together with other known USPCBs from \cite{2023AJ....165...80P} and new models 3a-c. The latter describes the evolution of a close binary with masses $0.89+0.46 \, M_{\sun}$ and initial period 2.2~d (model 3a), $1.01 +0.35 \, M_{\sun}$ and 2.5~d (model 3b), and $0.96+0.35 \, M_{\sun}$ and 2~d (model 3c), each evolved from the ZAMS until the present state. We assume a constant mass transfer rate after RLOF of the more massive component around $5.5 -7.7 \times 10^{-9} \, M_{\sun}~{\rm yr}^{-1}$,  and a lower rate of  $2.9 \times 10^{-10} \, M_{\sun}~{\rm yr}^{-1}$ during the contact phase (see a detailed discussion in \cite{2015A&A...577A.117S} and in \cite{2023AJ....165...80P}.  At each time step, all stellar parameters are interpolated from the PAdova and TRieste Stellar Evolution Code \citep[PARSEC;][]{2012MNRAS.427..127B} grid using the metallicity value of ${\rm [Fe/H]}=-1.58\pm 0.10$~dex for CRTS$\_$J163819 from {\em Gaia}'s DR3 \citep{GaiaDR3}. As seen in Figure~\ref{fig:Stepien}a, the progenitors best fitting the present parameters of CRTS$\_$J163819 are described with models 3a and 3b (cyan and orange lines, respectively). The model progenitors require around 10.7 and 7.2~Gyr, respectively, to match the current properties of the system. 
As discussed in \cite{2023AJ....165...80P}, the metallicities of USPCBs tend to be lower than found in other contact systems, revealing that they are an old population. 
Due to the absence of additional information (e.g. age), both solutions could be considered equally probable as progenitors. However, there is a weak indication that more extreme mass ratios come from binaries also with more extreme initial mass ratios. 

Under this scenario, we consider that the system must first reduce its period near the short limit when it reaches contact, after mass ratio reversal, and later achieves the low mass ratio while it evolves in contact, keeping a constant mass transfer rate. The opposite 
scenario, in which an LMR contact binary can evolve to a USPCB, does not seem to be plausible, as the mass transfer rate is lower during the contact phase and the latter's mean duration is $0.8 -1$~Gyr. Even if we consider AM exchange and tidal interaction with a distant tertiary companion, 
it is difficult for the Kozai-Lidov mechanism \citep{2007Fabrycky} to reduce dramatically the orbital period of an LMR system. 
We also display CRTS$\_$J163819 in the $P-J_{\rm s}/J_{\rm o}$ plane (Figure~\ref{fig:Stepien}b), together with the above compilation of USPCBs and the above models. It follows from the above approach that further theoretical research is urgently needed to address the accurate properties of the ancestors of cool contact EBs probably through hydrodynamical models and a proper stellar evolution code that includes the contact phase.  

\bibliography{bibliography}{}
\bibliographystyle{aasjournal}



\end{document}